\documentclass[twocolumn,twoside,slac_two]{revtex4}
\pdfoutput=1
\usepackage{graphicx}
\usepackage{fancyhdr}
\begin{document}

\title{Study of the Gamma-ray performance of the GAMMA-400 Calorimeter}

\author{P. Cumani}
\affiliation{Istituto Nazionale di Fisica Nucleare, Sezione di Trieste and Physics Department of University of Trieste, Trieste, Italy}
\author{A.M. Galper, N.P. Topchiev, V.A. Dogiel,	Yu.V. Gusakov, S.I. Suchkov}
\affiliation{Lebedev Physical Institute, Russian Academy of Sciences, Moscow, Russia}

\author{I.V. Arkhangelskaja, A.I. Arkhangelskiy,	G.L. Dedenko, V.V. Kadilin, V.A. Kaplin, A.A. Kaplun, M.D. Kheymits, A.A. Leonov, V.A. Loginov, V.V. Mikhailov,	P.Yu. Naumov, M.F. Runtso, A.A. Taraskin, E.M. Tyurin,	 Yu.T. Yurkin, V.G. Zverev}
\affiliation{National Research Nuclear University MEPhI, Moscow, Russia}

\author{S.G. Bobkov, M.S. Gorbunov, A.V. Popov, O.V. Serdin}
\affiliation{Scientific Research Institute for System Analysis, Russian Academy of Sciences, Moscow, Russia}

\author{R.L. Aptekar, E.A. Bogomolov, M.V. Ulanov, G.I. Vasilyev}
\affiliation{Ioffe Institute, Russian Academy of Sciences, St. Petersburg, Russia}

\author{A.L. Men'shenin}
\affiliation{Research Institute for Electromechanics, Istra, Moscow region, Russia}

\author{V.N. Zarikashvili}
\affiliation{Pushkov Institute of Terrestrial Magnetism, Ionosphere, and Radiowave Propagation, Troitsk, Moscow region, Russia}

\author{V. Bonvicini, M. Boezio, F. Longo, E. Mocchiutti, A. Vacchi, N. Zampa}
\affiliation{Istituto Nazionale di Fisica Nucleare, Sezione di Trieste and Physics Department of University of Trieste, Trieste, Italy}

\author{R. Sarkar}
\affiliation{Indian Centre for Space Physics, Kolkata, India}

\author{O. Adriani, E. Berti, M. Bongi, S. Bottai, N. Mori, P. Papini, P. Spillantini, A. Tiberio, E. Vannuccini}
\affiliation{Istituto Nazionale di Fisica Nucleare, Sezione di Firenze and Physics Department of University of Florence, Firenze, Italy}

\author{G. Castellini, S. Ricciarini}
\affiliation{Istituto di Fisica Applicata Nello Carrara - CNR and Istituto Nazionale di Fisica Nucleare, Sezione di Firenze, Firenze, Italy}

\author{G. Bigongiari, S. Bonechi, P. Maestro, P.S. Marrocchesi}
\affiliation{Department of Physical Sciences, Earth and Environment, University of Siena and Istituto Nazionale di Fisica Nucleare, Sezione di Pisa, Italy}

\author{P.W. Cattaneo, A. Rappoldi}
\affiliation{Istituto Nazionale di Fisica Nucleare, Sezione di Pavia, Pavia, Italy}

\author{I. Donnarumma, G. Piano, S. Sabatini, V. Vittorini}
\affiliation{Istituto di Astrofisica e Planetologia Spaziali, Rome, Italy}

\author{A. Argan}
\affiliation{Istituto Nazionale di Astrofisica, Rome, Italy}

\author{A. Bulgarelli, V. Fioretti}
\affiliation{Istituto di Nazionale di Astrofisica Spaziale e Fisica Cosmica, Bologna, Italy}

\author{M. Tavani}
\affiliation{Istituto di Astrofisica e Planetologia Spaziali and University of Rome Tor Vergata, Rome, Italy}

\author{C. De Donato, P. Picozza, R. Sparvoli, F. Palma}
\affiliation{Istituto Nazionale di Fisica Nucleare, Sezione di Roma 2 and Physics Department of University of Rome Tor Vergata, Rome, Italy}

\author{M. Tavani}
\affiliation{Istituto Nazionale di Astrofisica IASF and Physics Department of University of Rome Tor Vergata, Rome, Italy}

\author{L. Bergstr{\"o}m}
\affiliation{Stockholm University, Department of Physics; and the Oskar Klein Centre, AlbaNova University Center, Stockholm, Sweden}

\author{	J. Larsson, M. Pearce, F. Ryde}
\affiliation{KTH Royal Institute of Technology, Department of Physics; and the Oskar Klein Centre, AlbaNova University Center, Stockholm, Sweden}

\author{A.A. Moiseev}
\affiliation{CRESST/GSFC and University of Maryland, College Park, Maryland, USA}

\author{I.V. Moskalenko}
\affiliation{Hansen Experimental Physics Laboratory and Kavli Institute for Particle Astrophysics and Cosmology, Stanford University, Stanford, USA}

\author{B.I. Hnatyk}
\affiliation{Taras Shevchenko National University of Kyiv, Kyiv, Ukraine}
\author{V.E. Korepanov}
\affiliation{Lviv Center of Institute of Space Research, Lviv, Ukraine}

\begin{abstract}
GAMMA-400 is a new space mission, designed as a dual experiment, capable to study both high energy gamma rays (from $\sim$100 MeV to few TeV) and cosmic rays (electrons up to 20 TeV and nuclei up to $\sim$10$^{15}$ eV). The full simulation framework of GAMMA-400 is based on the Geant4 toolkit. The details of the gamma-ray reconstruction pipeline in the pre-shower and calorimeter will be outlined. The performance of GAMMA-400 (PSF, effective area) have been obtained using this framework. The most updated results on them will be shown.

\end{abstract}

\maketitle

\thispagestyle{fancy}
\section{INTRODUCTION}
GAMMA-400 (\cite{2013arXiv1306.6175G}) is a Russian space mission, approved by the Russian space agency, with an international contribution. Foreseen to be launched at the beginning of the next decade, the satellite will be positioned on a circular orbit at $\sim$200000 km. This particular orbit, combined with a pointing mode observational strategy, permits to perform continuous observations of a source without Earth occultation. During its first year of operation, GAMMA-400 is planned to observe the Galactic plane.\\
Designed as a dual experiment, GAMMA-400 will be able to study gamma rays, from 100 MeV up to several TeV, as well as cosmic rays, electrons up to 20 TeV and protons and nuclei up to the knee (10$^{15}$-10$^{16}$ eV). It will search for possible dark matter signal thanks to an unprecedented energy resolution that will permit to detect features associated to dark matter annihilation or decay in the spectra of sources such as the Galactic Center. GAMMA-400 will also study gamma-ray sources such as active galactic nuclei, supernova remnants, pulsars and gamma-ray bursts (GRBs).\\ 
GAMMA-400 will address the remaining issues regarding cosmic-rays origin, acceleration and propagation by studying the high energy all electron spectrum, with a 2\% energy resolution, and the cosmic-ray elemental spectra up to the knee, with high statistics and energy resolution.\\
A gamma-ray reconstruction pipeline using only the pre-shower or the calorimeter will be presented in the following. The reconstruction pipeline is part of a larger framework based on the Geant4 (\cite{Agostinelli:2002hh}) toolkit. The framework contains tools to create the geometry, simulate the particle interactions inside the apparatus, digitize the output of the simulations and analyze the results as well as an event displayer.

\section{GEOMETRY}

The GAMMA-400 apparatus, of which a schematic view is presented in fig. \ref{baseline}, will comprise:
\begin{figure}[h]
\centering
\includegraphics[width=0.43\textwidth]{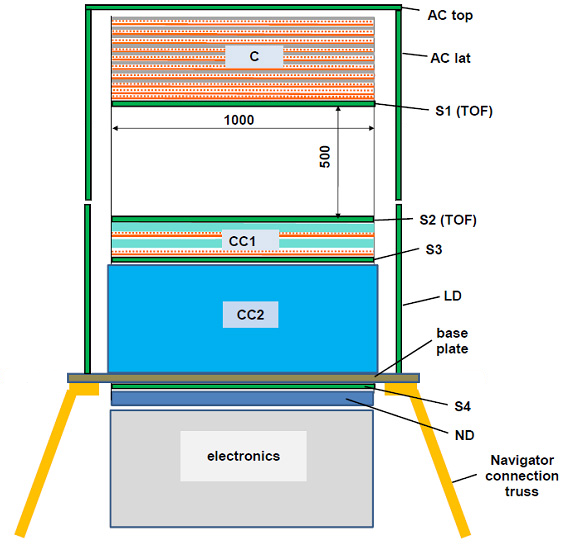}
\caption{GAMMA-400 physical scheme. The dimensions values are in mm.} \label{baseline}
\end{figure}

\begin{figure*}[t]
\centering
\includegraphics[width=0.49\textwidth]{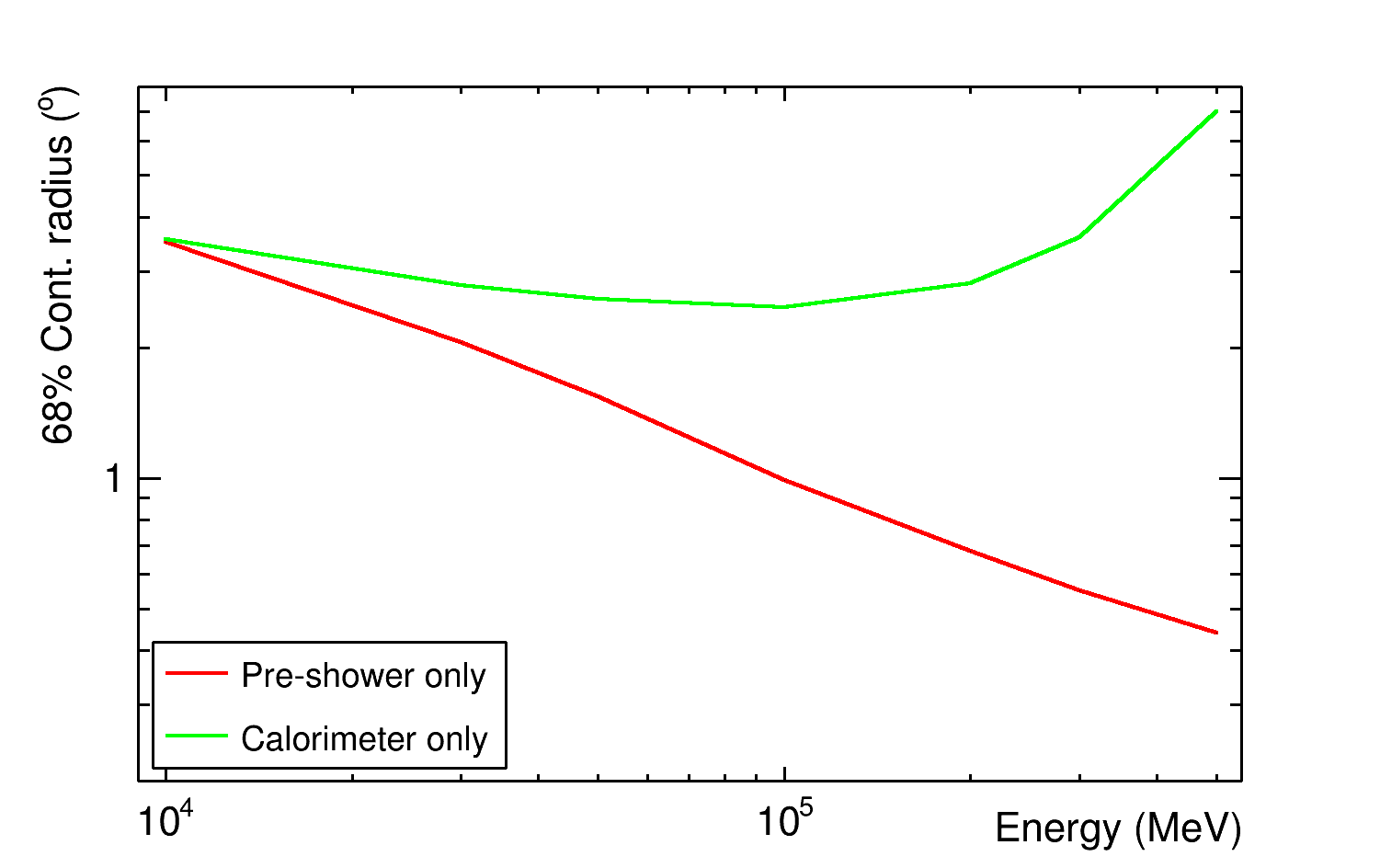}
\includegraphics[width=0.49\textwidth]{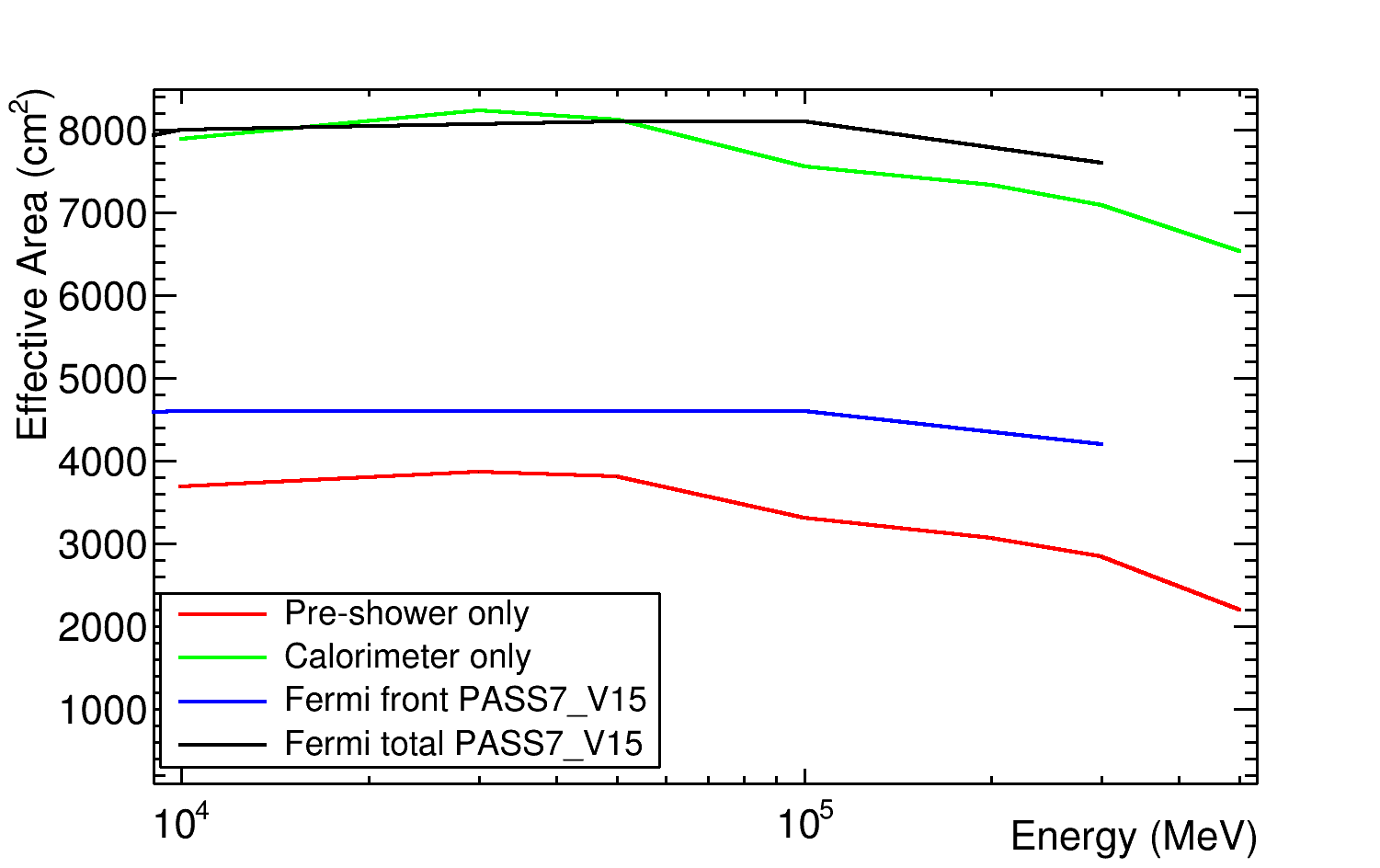}
\caption{Angular resolution (\textit{left}) and effective area (\textit{right}) of GAMMA-400, using its different detectors. The results on the effective area are compared with the performance of \textit{Fermi}-LAT (\cite{URL})}
\label{perf}
\end{figure*}
\begin{itemize}
\item A converter/tracker (C) where a gamma ray interacts with a tungsten layer (8 layers, 0.1 X$_0$ each) creates an electron-positron pair subsequently detected by single-sided Silicon layers (10 layers);
\item A calorimeter composed partially by two planes of CsI(Tl) slabs and Silicon (CC1, also referred to as pre-shower in the following) and partly by CsI(Tl) cubes (CC2, also referred to as calorimeter in the following), arranged in a 28$\times$28$\times$12 array;
\item An Anticoincidence system covering both the sides and the top of the detector (AC top and lat) to reject the charged particles for gamma-ray observations. The possibility to retrieve also timing information from the AC is currently under study;
\item A Time-of-flight system composed by four layers of scintillating materials (S1 and S2) to discriminate upgoing particles, such as backsplashed particles from the calorimeter, and downgoing particles;
\item A charge identification system (LD), to discriminate between the different elements interacting inside the detector;
\item A neutron detector (ND) and scintillation detectors (S3 and S4), used to improve the electron/hadron rejection factor.
\end{itemize}

\section{DIRECTION RECONSTRUCTION}
The reconstruction of the direction of an incoming gamma ray with the GAMMA-400 apparatus can be performed using either the combination of information from the tracker, pre-shower and calorimeter or only one of these detectors. While the results of the reconstruction using also information from the tracker are presented in, e.g., \cite{2014arXiv1412.4239G}, the reconstruction using only the pre-shower or the calorimeter will be discussed in the following.

\subsection{Pre-shower Only Reconstruction}\label{s:presonly}
A direction reconstruction can be performed using information from only the pre-shower. The requirement is for both Si planes to be hit. On each plane a median weighted on the energy is computed. The resulting points are fit through a straight line. The method is iterated several times, excluding the hits outside a cylinder centered along the found direction and reducing the cylinder radius at each iteration.

\subsection{Calorimeter Only Reconstruction}\label{s:caloonly}
Thanks to the novel configuration of the calorimeter, it is possible to reconstruct the shower created by particles coming not only from the top but also from the side of the detector as well as their original direction. The direction reconstruction method is similar to the one described in sec. \ref{s:presonly}, but it starts with a rough estimation of the original direction of the incoming particle. This estimation is needed to define the inclination of the planes on which the points to fit are computed. The planes are defined as perpendicular to the direction result of the fit of the three cubes with the highest energy release. Since only a rough estimation is needed, the inclination of the planes is rounded to the nearest $\pi /4$ multiple.\\
At least three hit planes are necessary for the reconstruction. No requirement on the containment of the shower are applied.
Once the planes are found, an average weighted on the energy is performed on each of them to find the barycenter. The barycenters on different planes are fit and the method is iterated by excluding the hit outside a cylinder centered on the found direction and reducing the cylinder radius after each iteration.\\

\subsection{Results}
The events that contributes to the calorimeter only and pre-shower only angular resolution and effective area are the events lacking of an overall reconstruction. The overall reconstruction, making use of the tracker, is indeed better and these other events are used to improve the effective area rather than the angular resolution. The sum of the three different effective areas gives an estimation of the total effective area of the instrument.\\
The angular resolution of the pre-shower improves with the energy because the identification of the hit in the pre-shower is made easier by the higher energy of the pair. Since no requirements on the containment of the shower in the calorimeter are applied, the angular resolution of the calorimeter only reconstruction decreases at high energy.\\
The different effective area between GAMMA-400 and \textit{Fermi} is due not only to the different reconstruction algorithm but also to the difference in the geometrical area of the two instrument.\\

\section{CONCLUSIONS}

GAMMA-400 is a dual experiment dedicated to the study of both gamma rays and cosmic rays, electrons, protons and nuclei. It is possible to reconstruct the direction of the incoming gamma ray using information not only from the tracker, as shown in \cite{2014arXiv1412.4239G}, but also from only the pre-shower or the calorimeter. The results of these type of reconstruction can be used to increase the total effective area of the instrument, at the expense of the angular resolution. The calorimeter, because of a novel configuration, is capable of reconstruct the direction of particles coming also from its sides, resulting in a more than $2\pi$ sr field-of-view. It can then be used to provide a trigger for observations of transients from the ground with telescopes such as the future CTA.\\

\bigskip 
\bibliography{biblio}

\end{document}